\documentclass[preprint, aps,a4paper,showkeys,showpacs]{revtex4-1}

\usepackage{amsmath}\usepackage[subrefformat=parens]{subcaption}
\usepackage{xcolor}

\newsavebox{\boxa}
\newsavebox{\boxb}

\makeatletter
\def\btt#1{\texttt{\@backslashchar#1}}%
\DeclareRobustCommand\bblash{\btt{\@backslashchar}}%
\makeatother

\usepackage{booktabs}
\usepackage{multirow}
\usepackage{rotating}

\RequirePackage[l2tabu, orthodox]{nag}

\newcommand{\figref}[1]{Fig. \ref{#1}}

\graphicspath{{figs/}}

\begin{document}

\preprint{1805.05598}

\title[Short Title]{Diverse Stochasticity Leads a Colony of Ants to Optimal Foraging \\}


\author{Masashi Shiraishi$^{1,3}$}

\author{Rito Takeuchi$^{4}$}%

\author{Hiroyuki Nakagawa$^2$}%


\author{Shin I Nishimura$^5$}%

\author{Akinori Awazu$^1$}%

\author{Hiraku Nishimori$^1$}%
\email{nishimor@hiroshima-u.ac.jp}
\affiliation{${}^1$Department of Mathematical and Life Sciences, Hiroshima University, Kagamiyama, Higashi-hiroshima 739-8526, Japan
}%
\affiliation{${}^2$Department of Mathematical Sciences, Osaka Prefecture University, Sakai 599-8531, Japan
}%
\affiliation{${}^3$CREST, JST, Tokyo, Japan
}%
\affiliation{${}^4$Yahoo Japan Corporation, Tokyo, Japan
}%
\affiliation{${}^5$JamGuard Corporation, 1-5-6 Kudam-Minami, Chiyoda, Tokyo, Japan
}

\date{\today}
 
\begin{abstract}

A mathematical model of garden ants (\textit{Lasius japonicus}) is introduced herein to investigate the relationship between the distribution of the degree of stochasticity in following pheromone trails and the group foraging efficiency.
Numerical simulations of the model indicate that depending on the systematic change of the feeding environment, the optimal distribution of stochasticity shifts from a mixture of almost deterministic and mildly stochastic ants to a contrasted mixture of almost deterministic ants and highly stochastic ants.
In addition, the interaction between the stochasticity and the pheromone path regulates the dynamics of the foraging efficiency optimization.
Stochasticity could strengthen the collective efficiency when the variance in the sensitivity to pheromone for ants is introduced in the model.
\end{abstract}


\maketitle

\section{Introduction}
Ants efficiently shuttle between their nest and food sources over long distances by depositing and following pheromone trails\cite{Wil}.
As a result of the collective usage of pheromones, ants are able to construct a trail that might be accomplished by accurate pheromone usage.
Following the trail ensures that ants can accurately reach food sources.
However, a certain fraction of ants in each colony does not accurately follow pheromone trails\cite{De0,Evi}.
The concept of \textit{stochastic foraging} or \textit{strategy of error} is introduced and validated through experiments and mathematical models to interpret the ecological significance of this stochasticity\cite{De0,Dus,Nic,Cal1,Cal2,Couzin,Verha,Bo,Schw,Do}. 
Deneubourg et al. performed numerical experiments to understand the meaning of the very low percentage of \textit{Tetramorium impurum} reaching a food source following a pheromone trail\cite{De0}.
Using a simple phenomenological model for the recruiting dynamics of ants, they estimated the optimized degree of stochasticity for each ant to be recruited to an already-found food source in a multi-food-source environment.
They found that a less accurate recruitment led the colony to a more efficient foraging as the number of location of food sources increased.
These studies recognized the stochasticity of ants in following pheromone trails as a factor that increased their chance to find new food sources and avoid excessive persistence to previously established trails.
Nowadays, the concept of ``strategy of error" is also known as ``explore--exploit" trade-off problem in diverse research fields.
Studies share the similar structures and solutions, even though research subjects are different from each other\cite{Hills46-54}.

However, many basic problems on stochasticity-induced efficient foraging remain unsolved.
One among them is the main issue being focused on herein: what type of population distribution of stochasticity enables optimized foraging in a colony of ants? 
Previous models of stochastic foraging have implicitly or explicitly introduced the uniform population distribution of stochasticity because of their simple forms\cite{De0}.
However, experiments with existing ants have shown widely dispersed population distributions of stochasticity.
As an extreme example, only 18$\%$ of \textit{Tetramortium inpurum} in a colony can reach the food sources by following pheromone trails, and the same quantity decreases down to 9$\%$ in the case without a leader\cite{De0}.
As another example, typically more than $40\%$ individuals in a colony of garden ants species \textit{Lasius niger}, which usually make use of both visual and chemical cues in their foraging trip, fail to follow pheromone trails if visual cues are made imperfectly available\cite{Evi}.
We believe that the population distribution of stochasticity in following pheromone trails is not always restricted to a Gaussian-like unimodal form.
We evaluate this presumption by performing a model-based study on what type of distribution of stochasticity leads a colony of ants to efficient foraging.
We focus on the effects of the intensity of stochasticity of the stochastic ants and their population rate in the colony under different environment conditions.
Our study aims to mathematically understand how the stochasticity in a complex collective system based on real biological systems works to optimize the collective behaviors and analyze how the specific stochasticity in the ant colony model works compared to the stochasticity in previous studies\cite{De0,Dus,Nic,Cal1,Cal2,Couzin,Verha,Bo,Schw,Do}.

This paper is organized as follows: Section 2 introduces our agent-based model; Section 3 explains the evaluation procedure of the foraging behaviors with the model; Section 4 shows the numerical results and discusses some combinations of stochasticity intensity and their population rate in the colony; and Section 5 concludes and summarizes the results.

\section{Model}
\begin{figure}
\begin{minipage}[b]{0.45\linewidth}
\centering
\includegraphics[scale=0.3]{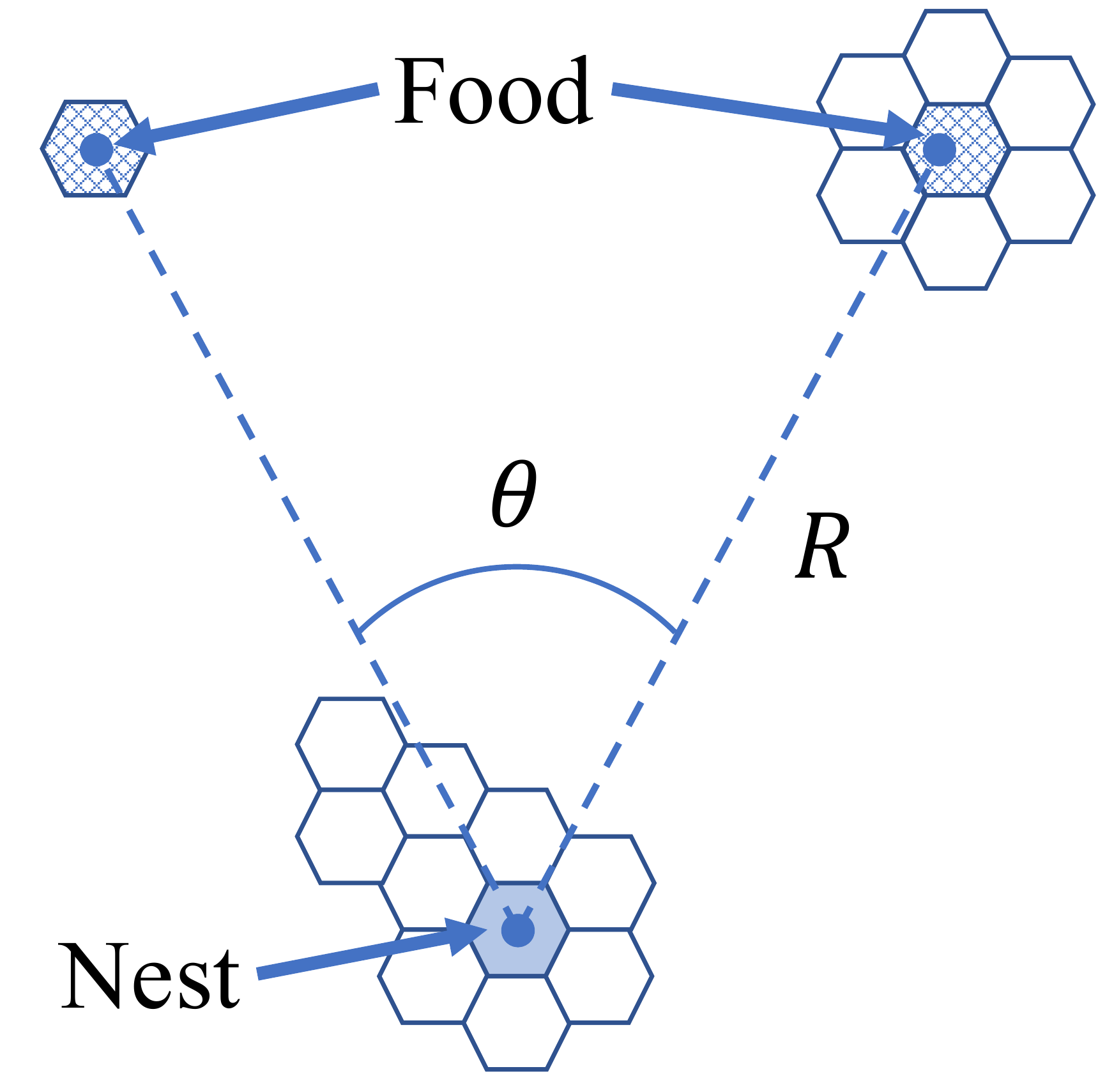}
\subcaption{}
\end{minipage}
\begin{minipage}[b]{0.45\linewidth}
\centering
\includegraphics[scale=0.3]{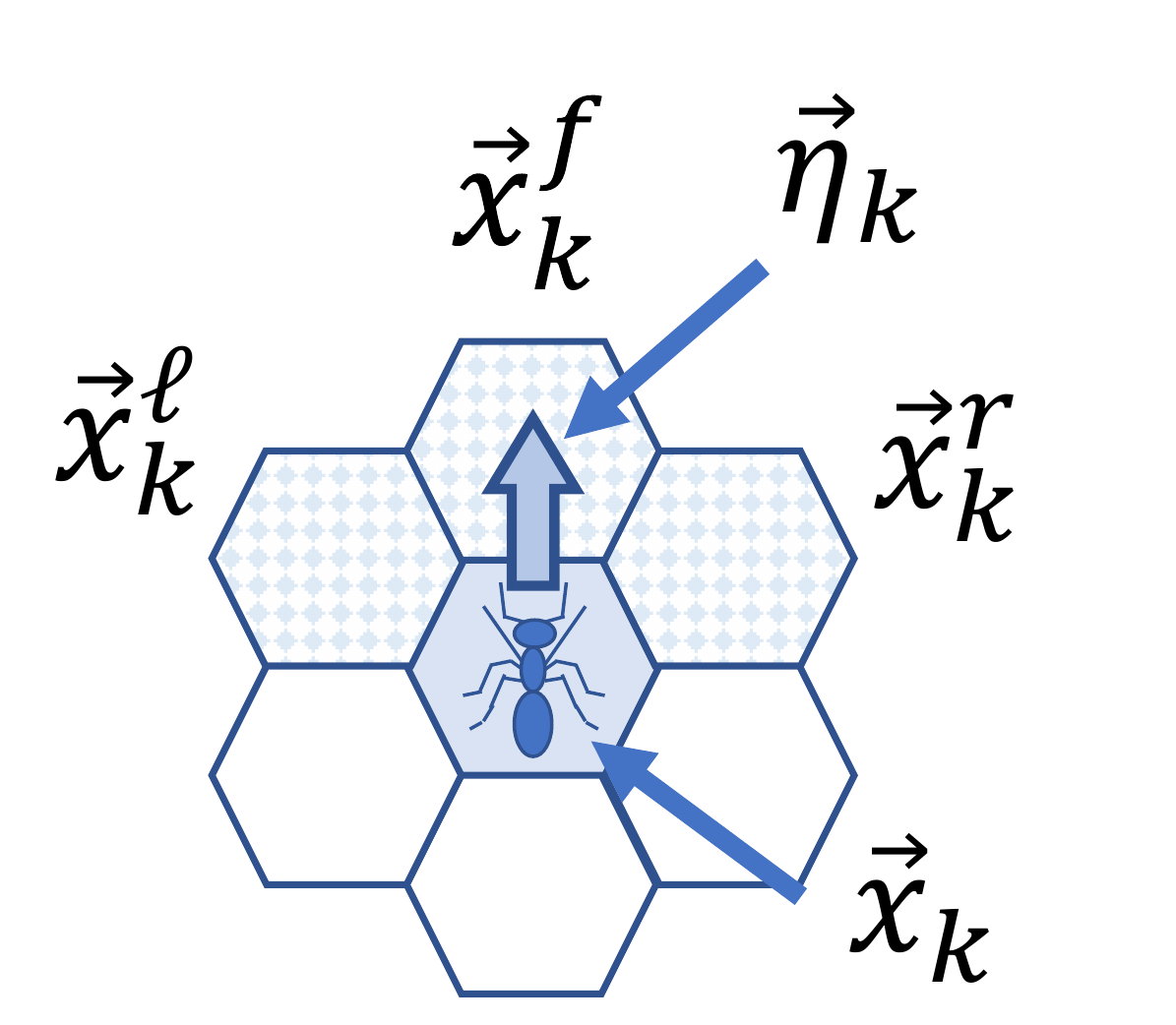}
\subcaption{}
\end{minipage}
\caption{(a) The distance $R$ between the nest and the food source points and the relative angle $\theta$ between two food sources as seen from the nest are considered as the feeding environmental parameters. Food resource sites are the closest hexagonal sites (blue meshed hexagonal sites) to the food source points (blue dots). (b) Movable sites $\{\vec{x}_k^\ell, \vec{x}_k^f, \vec{x}_k^r\}$ of ants.}  \label{FoodSources}
\end{figure}

We employ herein a computational model categorized in multi-agent models \cite{code, Schw, Ogi}.
The unique task for each agent (i.e., ant) in the present model is to find food sources and bring food to the nest.
The peculiar feature of this model is that in addition to using trail pheromone, ants make use of visual cues to roughly grasp the homing direction, where this non-local walking rule of ants is introduced to make a qualitative correspondence to the behavior of the existing species of ants, namely the \textit{Lasius japonicus}.
The foraging field is a hexagonal two-dimensional lattice measuring 150 $\times$ 150 with a periodic boundary condition, on which one nest site and two sites of food sources are designated.
The dynamical states for each ant, which are labeled as $k\in \{1, 2, \dots N_{total}\}$, are represented by a positional vector, $\vec{x}_k$, consisting of a lattice index for the hexagonal two-dimensional lattice and a facing direction vector, $\vec{\eta}_k$, consisting of one of the unit vectors of the hexagonal two-dimensional lattice.
Each ant at each time step is located at one site, but does not exclusively occupy it, and faces toward one of its nearest six sites.
A randomly selected single ant among a total of $N_{total}$ ants moves to the facing site, its right site, or its left site (the detailed process to select a site is explained below).
The same processes are repeated for $N_{total}$ times and constituted a unit Monte Carlo (MC) step.
This unit MC step is accompanied by a single step of a difference equation used to update the amount of food and another difference equation used for the evolution of the pheromone field, which will be explained in Sections 2F and C, respectively.
The distance $R$ between the nest and the food source sites and the relative angle $\theta$ between two food source sites as seen from the nest are given as the feeding environmental parameters~(\figref{FoodSources}).
Increasing the food supplies per unit time is also an important feeding environment parameter, which characterizes the dynamical aspects of this model.
The dynamics of food supply and the evaluation methods of the system are explained in the sub-sections that follow.

\subsection{Chemotactic Dynamics}
In this model, each ant moves to the facing site or its adjacent sites, which are stochastically decided upon depending on the trail pheromone on the lattice field.
$\vec{x}_k$ is the position of the $k$th ant, and $\vec{\eta}_k$ is its facing direction vector.

\begin{figure*}
\centering
\includegraphics[width=\textwidth ]{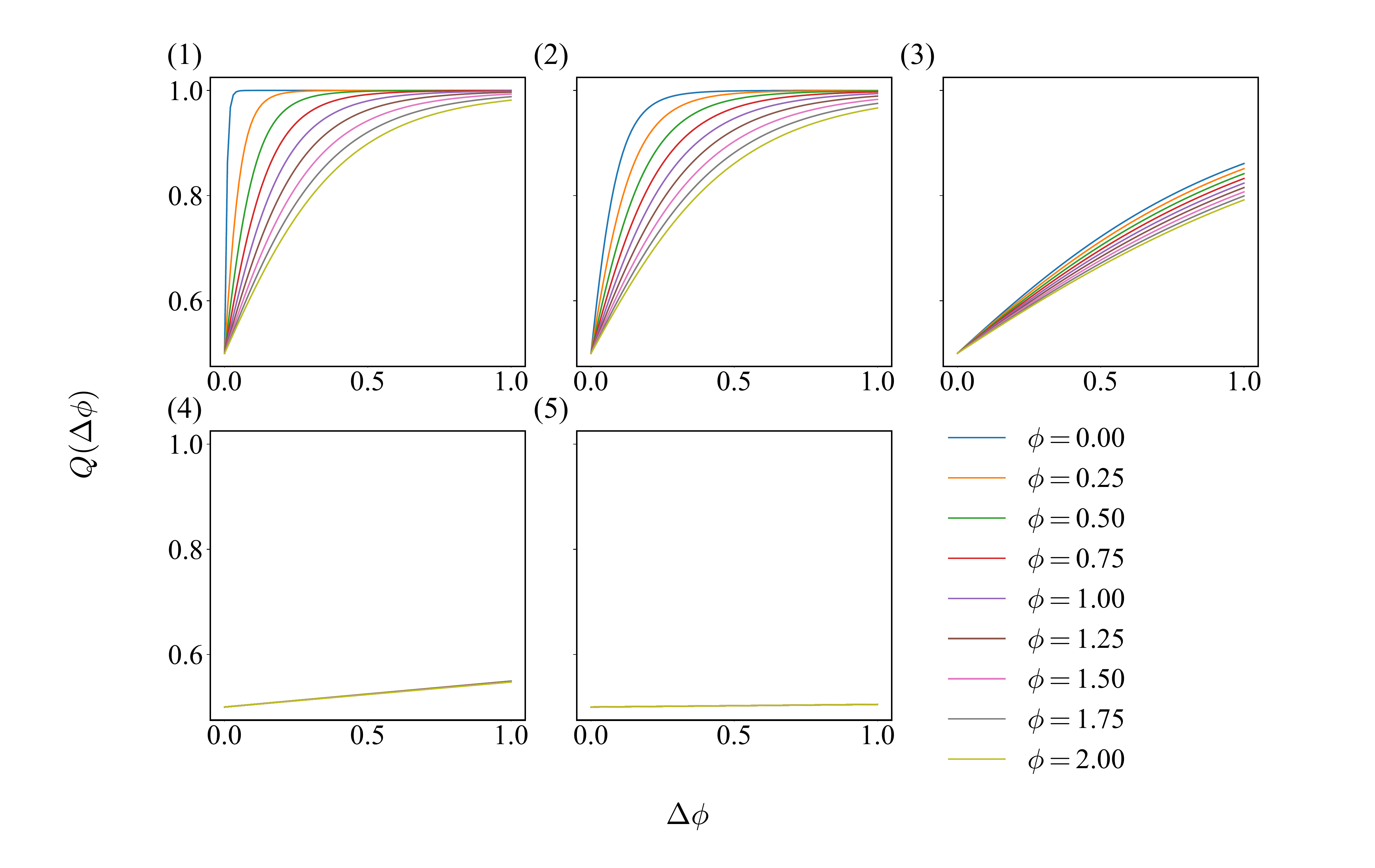}
\caption{The pheromone density difference, $\Delta \phi$, vs the probability of choosing the cell with higher density between two cells, $Q(\Delta \phi) = (\alpha(\phi_0+\Delta\phi)+z)^h/(\alpha\phi_0+z)^h + (\alpha(\phi_0+\Delta\phi)+z)^h$, are shown, where $h=10$, $z=0.05$, and $\phi_0$ is the lower pheromone density of the other cell. $Q(\Delta \phi)$ is a simplified function of Eq. \eqref{Eq1}. The sensitivity parameters are: (1) $-\log_{10} \alpha_s = 0$, (2) $-\log_{10} \alpha_s = 1$, (3) $-\log_{10} \alpha_s = 2$, (4) $-\log_{10} \alpha_s = 3$, and (5) $-\log_{10} \alpha_s = 4$. (1) represents "normal" ants, while (2)--(5) represent "stochastic" ants with different parameters.
}  \label{ProbDiff}
\end{figure*}

Therefore, the site to which the ant faces is represented by $\vec{x}_k^f=\vec{x}_k+\vec{\eta}_k$, and $\vec{x}_k^\ell$ and $\vec{x}_k^r$ are the left and the right sites of the facing site, respectively.
The response probability, $p_j$, denoting the move to one of the three movable sites, $\vec{x}_k^j$ $(j\in \{\ell, f, r\})$, is presented as:
\begin{subequations}
\begin{align}
p_{j}(\phi(\vec{x}^\ell_k), \phi(\vec{x}^f_k), \phi(\vec{x}^r_k)) &= \frac{g\left( \phi\left(\vec{x}_k^j\right) \right) }{\sum_{{i}\in \{\ell, f, r\}}^{3} g\left(\phi\left(\vec{x}_k^{i}\right)\right)} \label {Eq1}, \\
g(y) &= \left(\alpha_k y + z\right) ^h \label{Eq2},
\end{align}
\end{subequations}
where $\phi\left(\vec{x}_k^j\right)$ is the trail pheromone density at $\vec{x}_k^j$; a parameter $h=10$ corresponds to the Hill coefficient, $z=0.05$; and $g(y)$ is a function defining the individual's susceptibility to the pheromone.
Unfortunately, this response function is not directly experimentally determined; thus, we adopted a function with a strong non-linear effect to make the response function of the pheromone a sigmoid-like function of $\phi(\vec{x}^\ell_k)$, $\phi(\vec{x}^f_k)$, and $\phi(\vec{x}^r_k)$.
Note that $\alpha_k$ is the key parameter controlling the stochasticity of the $k$th ant; hence, the sensitivity of the $k$th ant to the trail pheromone increases as $\alpha_k$ increases.
In this sense, we call $\alpha_k$ as the ``sensitivity parameter."
Figure \ref{ProbDiff} shows the effect of $\alpha_k$ on the probability to choose a higher density cell.
In the case of $-\log_{10}\alpha_k=4$, the probability remains at 0.5 even though $\Delta \phi$ increases (Fig. \ref{ProbDiff}(5)).
Aside from introducing the sensitivity parameter in this manner, the forms of Eqs. \eqref{Eq1} and \eqref{Eq2} follow those of the previous studies\cite{Nic,De2}.

We define ``normal ants," whose sensitivity parameter is fixed as $\alpha_k=\alpha_{n}=1$, as a criterial parameter.
We define ``stochastic ants," whose sensitivity parameter $\alpha_k =\alpha_{s}\in \{10^{-1}, \dots 10^{-4} \}$.
Note that the non-linear effect, which is determined by the value of $h$, controls the detectable range of pheromone concentration using the response function.
The detectable range affects the choice of $\alpha_k$ to determine the stochasticity of the ant behaviors.
According to the previous studies\cite{Bo}, the response function works similarly if $h > 1$.
If we choose a small value of $h$, like $h \approx 2$, we have to choose a much wider range of value of $\alpha$ to distinguish ``normal ants"  from ``stochastic ants."
Therefore, we choose a relatively higher value $h=10$ to restrict the range of $\alpha_k$ as the simulation environment.

The facing direction of the ant is reset to the moving direction once the moving direction is selected.
It then moves to the facing cell.
After this movement, the facing directions are kept unchanged until the next move. 

\subsection{Three Behavioral Modes}
We introduce the three following behavioral modes to incorporate the situation-dependent behavior of ants: exploring mode (EM), homing mode (HM), and emergency homing mode (EHM).
The details of each mode are explained below.

\noindent
(i) \textit{Exploring mode}: 

Every ant initially leaves the nest in this mode and explores for food by tracing the trail pheromone deposited by its nest-mates (or the ant itself) in HM.
In other words, the ant's movement is based on simply stochastic dynamics.
This mode lasts until the ant finds food or becomes exhausted without finding food after a finite time, $t_{th}=150$, has passed since it left the nest. 
A randomly selected ant in each walking step moves into one of its facing sites or its neighboring two sites based on the stochastic dynamics explained in the previous section.
Upon reaching one of the six neighboring sites of food sources, if the temporal amount of food exceeds the threshold value $f_{th}=0.1$, the ant takes a certain amount of food $f_e (\le f_{th})$ from the site, then its mode shifts from EM to HM.

\noindent
 (ii) \textit{Homing mode}:
 
In HM, the ant tries to return to the nest with the deposited trail pheromone to recruit nest-mates. 
As a peculiar feature of the walking manner in HM, a combined use of visual and chemical cues is considered assuming the homing manner of \textit{Lasius japonicus}\cite{Ogi}.
Accordingly, visual cues are preferentially relied on. 
We assume that homing ants roughly ``recognize”  the nest direction by visual cue, for which they make use of landmarks. 
More specifically, viewing from a homing ant, if the relative angle between the nest direction and its facing direction exceeds $60^{\circ}$ irrespective of the local pheromone field, the ant turns to the right- or left-facing direction to decrease the relative angle to the nest, then walks to one of the newly faced sites, in which it will keep its facing direction until the next move. 
Alternatively, when the relative direction is less than or equal to $60^{\circ}$, the ant determines its walking direction depending only on the pheromone densities among its three facing sites according to Eqs. \eqref{Eq1} and \eqref{Eq2}.
In each walking step in HM, the ant deposits a fixed amount of trail pheromone, $\phi_{u}=0.08$, at site $x_i$, from which it leaves.

\noindent
(iii) \textit{Emergency homing mode}: 

If an ant in EM cannot find food within a certain time interval $t_{th}$ after leaving the nest or in HM and fails to come back to the nest within the same interval $t_{th}$ after getting food, its mode changes to EHM.
Aside from not depositing the pheromone, an ant in this mode moves in the same way as in the usual HM.
Note that we add EHM because of the limitation of our simulation to assume the steady state.
In the case that the system does not have EHM, the ant, which could not find food, would walk away from the simulation lattice space.
Consequently, the number of ants in the system decreases, and we would not obtain steady state data to calculate the statistical quantities.
Therefore, we define $t_{th} = 150$, which corresponds to the lattice size, to confine the ants in the simulation lattice space.
Moreover, the sensitivity parameter is set high as $\alpha_k =\alpha_{n}=1$ for every ant in this mode to enable the ant a prompt return to the nest after losing its way.
Upon arriving at one of the neighboring sites of the nest, the ant returns to EM. 

\subsection{Evolution of the Pheromone Field}
\begin{figure*}
\begin{minipage}[b]{0.3\textwidth}
\centering
\includegraphics[width=1.2\textwidth]{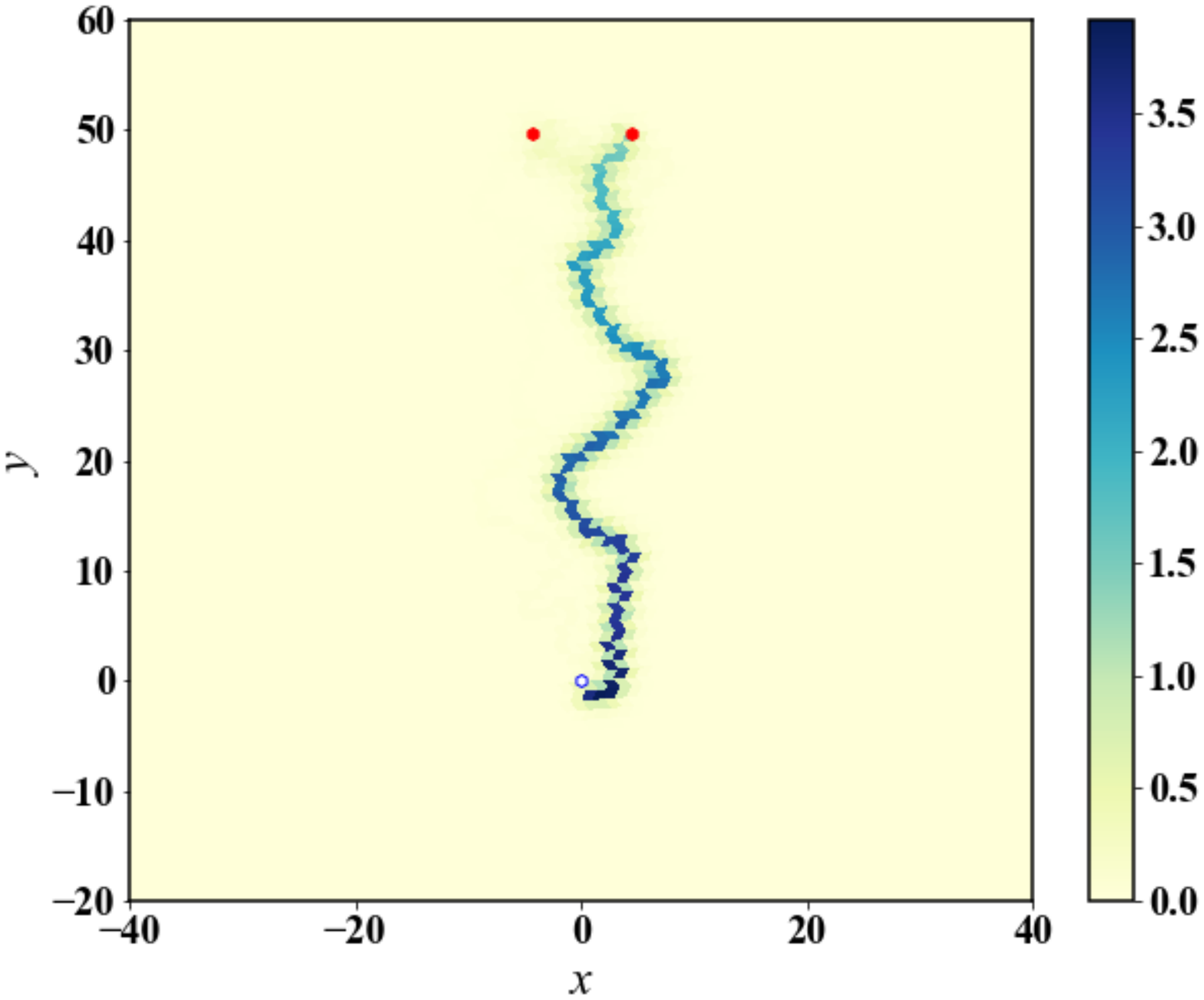}
\subcaption{$\delta=0.1$}
\end{minipage}
\begin{minipage}[b]{0.3\textwidth}
\centering
\includegraphics[width=1.2\textwidth]{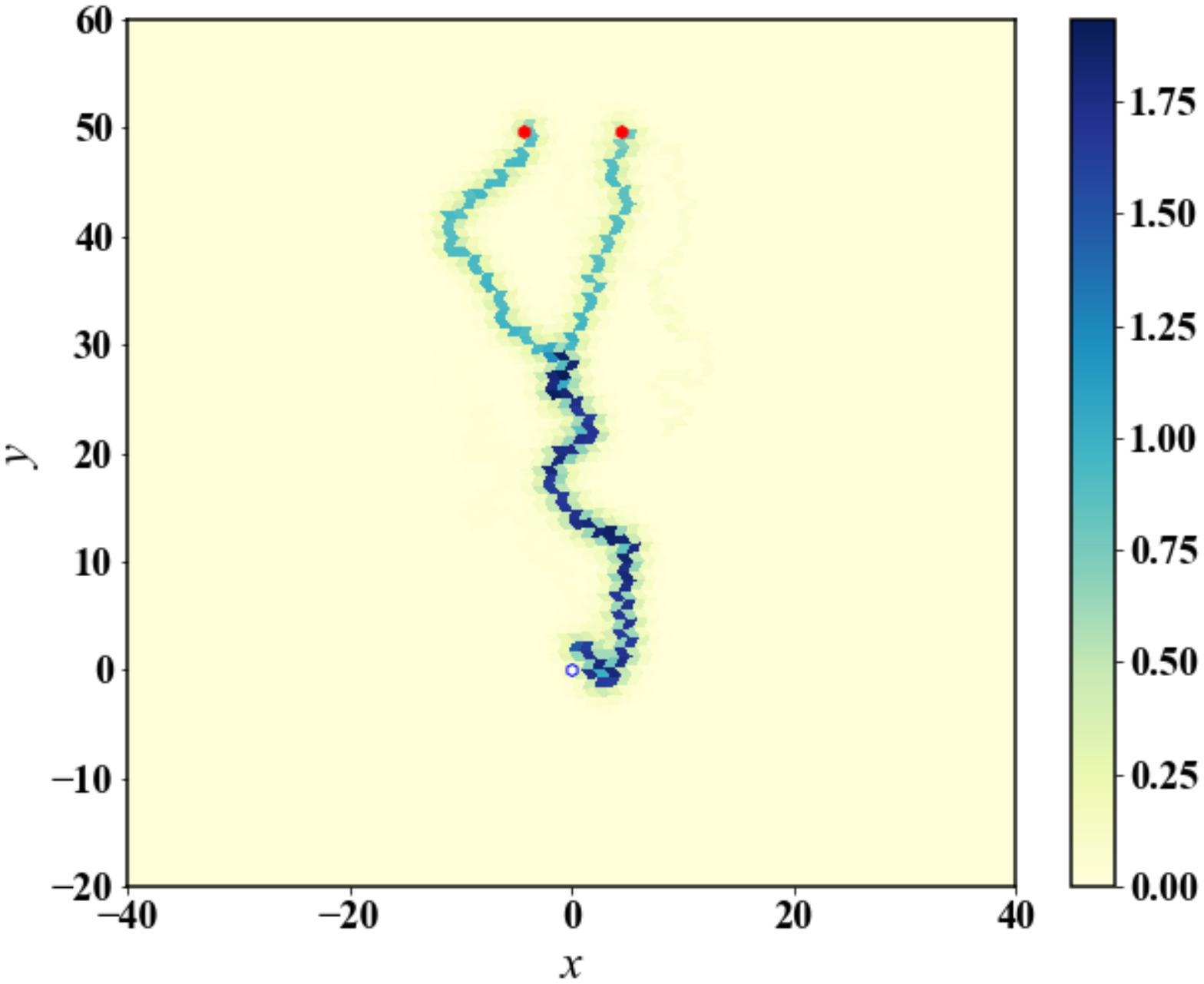}
\subcaption{$\delta=0.8$}
\end{minipage}
\begin{minipage}[b]{0.3\textwidth}
\centering
\includegraphics[width=1.2\textwidth]{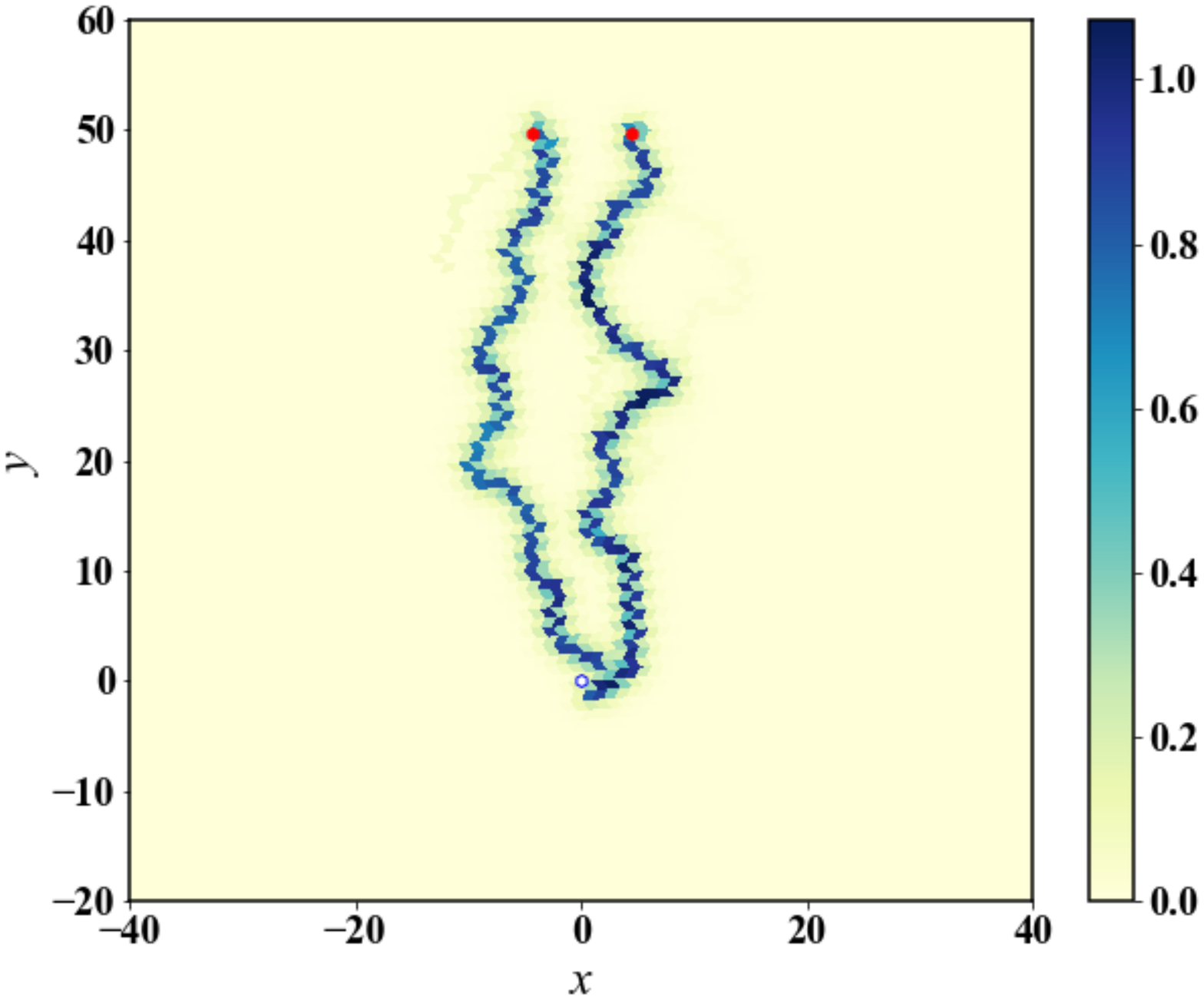}
\subcaption{$\delta=2.0$}
\end{minipage}
\caption{Snapshots of pheromone trails with $-\log_{10}\alpha=3$, $\hat{N}_n=0.5$, and different $\delta$. The white hexagonal box represents the nest position, while the red hexagonal boxes are the two food sources. (a) $\delta=0.1$. (b) $\delta=0.8$. (c) $\delta=0.1$.
}  \label{fig:samplepath}
\end{figure*}

The pheromone density $\phi(x,t)$ in the hexagonal sites $x$ is updated using the following equation: 

\begin{subequations}
\begin{align}
\label{eq:phero}
\phi^{t+\tau}(\vec{x}) &= \phi^{\textit{t}}(\vec{x}) - D_p\delta \phi^t(\vec{x}) - e_p\phi^t(\vec{x}) + n_h^t(\vec{x})\phi_u, \\
\delta \phi^t(\vec{x}) &= \phi^t(\vec{x})-\sum_{i=1}^{6} \frac{\phi(\vec{x}^i,t)}{6}
\end{align}
\end{subequations}
where $\tau$ is the unit time of the present model; the second term in the r.h.s. represents the discretized form of diffusion, in which $\delta \phi^t(\vec{x})$ is the difference from average around the nearest sites of site $\vec{x}$ as shown in the second equation; and $D_p$ is a diffusion constant.
The third term and the last term represent evaporation and the total amount of deposited pheromone at $\vec{x}$ during one MC step, respectively, where $n_{h}(\vec{x})$ is the number of times the homing ants leave the site $\vec{x}$ depositing pheromone during one MC step, and constants $e_p$ and $\phi_{u}$ are the evaporation rate of the pheromone and the amount of pheromone secreted by an ant, respectively.
Figure \ref{fig:samplepath} includes the snapshots of pheromone trails for different values of feeding increment of foods, $\delta$, which will be explained in the later subsection.

In our previous study \cite{Ogi}, we introduced two types of pheromone concentration fields, $p_g$ and $p_a$, to illustrate a more detailed dynamics of ant behaviors.
$p_g$ represents the pheromone concentration field, which is deposited by ants on the ground, and $p_a$ is the pheromone concentration field evaporated to air from the ground.
However, the time-scale of evaporation from the ground is sufficiently faster than that of diffusion of $p_g$.
Hence, we only consider herein the time development of the pheromone in the air.

\subsection{System Parameters}
As mentioned earlier, the present model assumes \textit{Lasius japonicus} as the corresponding species of ants; therefore, the parameter values are consistent with the assessable data of this ant species.
Time and space units are also set accordingly.
\begin{table}[htbp]
	\caption{System Parameters}
	\centering
	\begin{tabular}{lccc}
		\hline
		\  & Description & Amount & Unit \\
		\hline
		\hline
		\ $l_u$\ \ & Unit length & 1.5 & cm \\ 
        \ $\tau$\ \ & Unit time & 3 & sec \\ 
        \ $\phi_{u}$\ \ & Secreted pheromone amount & 0.08 & $\tau^{-1}$ \\
		\ $e_p$\ \ & Evaporation rate & 0.01 & $\tau^{-1}$ \\
		\ $D_p$\  \ & Diffusion constant & 0.015 & $l_u^2 \tau^{-1}$ \\
        \ $h$\ \ & Hill's coefficient &10 & \\
        \hline
	\end{tabular}
	\label{default}
\end{table}%

\noindent
First, the unit length, which is the distance between the neighboring cells, is set as approximately 1.5 cm, whereas the unit time, one MC step, $\tau$ is set as 3 s. 
Considering that each ant in the unit MC simulation walks once in average, the abovementioned estimations are consistent with the typical walking velocity, $\sim$ 0.5 cm/s, of \textit{Lasius japonicus}.
In addition, the evaporation rate of pheromone $e_p=0.01$ in this model is consistent with the experimentally obtained duration time (i.e., 5 min) of the pheromone.
Furthermore, the width of foraging trail is within a few centimeters.
Therefore, the characteristic width, $\sqrt{D_p/e_p} \sim1.2$ assuming $D_p=0.015$ in our model, is consistent with the reality.
We know that the preceding experiments~\cite{Bec} of \textit{Lasius niger} estimated 47~min as the lifetime, and those of Monomorius pharaohs~\cite{Jean} estimated the decay time (to 50\%) on the plastic and paper substrates to yield 25~min and 8~min, respectively. 
We would like to stress again that the evaporation rate (or life time) of the pheromone significantly depends on the species and the experimental situations.
Meanwhile, the Hill coefficient (n=2 in ref~\cite{De2,Nic}) is ``manually" set as $h=10$, such that the combination of the walking probability in the three facing directions sharply varies between $(1/3, 1/3, 1/3)$ and one of $\{(0,0,1), (0,1,0), (1,0,0)\}$ according to the change of $\alpha_k$ in the present range of its variation, thereby accentuating the effect of the non-uniformity of the stochasticity in a colony.

\subsection{Distribution of Sensitivity Parameters}
The basic motivation of this study is to investigate what type of distribution of stochasticity strength in a colony of ants will lead the colony to optimal foraging.
As the index controlling the distribution of stochastic ants in each colony, a set of sensitivity parameters $\{\alpha_k\},(k\in {1, \dots, N_{total}})$ introduced in Eq. \eqref{Eq2} is considered.
$\alpha_k$ of an ant generally has a different value corresponding to the individual variability of pheromone sensitivity; thus, the distribution of $\alpha_k$ would be continuous distributions.
However, we restrict the distributions to binary ones, the combination of ``normal ants" and ``stochastic ants", for the sake of easily understanding how the difference of stochasticity strength among ants in a colony affects collective foraging behaviors.

Moreover, fixing the total number of ants in each colony as $N_{total}=N_n+N_{s}=500$, we vary $N_n$, which is the number of normal ants in the same colony.
The ratio $\hat{N}_n=N_n/N_{total}$ is essential; hence, we characterize a colony state using $\hat{N}_n$ hereafter instead of $N_n$.

\subsection{Feeding Environments}
In nature, aside from ants, other animals also have to spend time to find food and avoid the risk of starvation.
The longer animals explore outside their nests to find food, the higher their risks of being attacked by other animals or dying by accident.
Animals manage the balance between the potential risks surrounding them.
The difficulty of minimizing the total risks comes from uncertain food sources, which are randomly distributed and supply a fluctuating amount of food.
The efficiency of an ant colony under reasonable conditions, which represent many different natural environments, must be studied; however, performing a simulation for all the possible situations is difficult.

We control herein the spatial configurations of food sources and the dynamics of the amount of food of individual sources.
As we explained in Section 2, the number of food sources is limited to two as the most simplified situation in multiple food sources.
The dynamics of the amount of food is unpredictable and complex in nature; thus, its mathematical modeling requires some kind of simplification.
Therefore, we set up typical situations with controlling the feeding amount of food to understand the effects of stochasticity under the steady state and the response to dynamical perturbation.

We define the ``continuous feeding method" and the ``specialized feeding method" to simulate the two situations.

\subsubsection{Continuous Feeding Method}
In the simulation, the amount of food changes, such that the food at the two feeding sites is constantly supplied with a certain amount of $\delta$ and carried away by foragers reaching the sites.
Let $f_i(t)$ be the amount of food in the $i$-th feeding site at time $t$.
The time evolution of $f_i(t)$ is given by the following discrete equation:
\begin{equation}
\label{eq:food}
f_i(t+\tau) = \left\{ \begin{array}{lcl} f_i(t) - n^f_i(t) f_e + \delta & & \mbox{if } f_i(t) < f_c \\ f_c & & \mbox{otherwise} \end{array} \right.
\end{equation}
where $n^f_i(t)$ is the number of ants that succeeded in carrying away a certain amount of food $f_e$ from the food source $i$ in unit MC step; $\delta$ is the feeding increment of food; $f_i(0)=f_c$; and $f_c$ is the upper threshold for the amount of food introduced to represent that the food sources are finite in nature.
We set $f_c$ as 10 in the simulation.
We control the richness of the food environment by changing $\delta$.
Note that the time evolution of the amount of food at individual food sites in Eq. \eqref{eq:food} independently proceeds.
This feeding method corresponds to the insufficient feeding environment when $\delta$ is small and sufficient feeding environment when $\delta$ is large.

\subsubsection{Specialized Feeding Method}
In addition to the abovementioned feeding method, we adopt  a specialized feeding situation to study how ants manage to find an alternative feeding site when the already discovered food source at one feeding site runs out.
To do so, at the beginning, a fixed amount of food is set in a feeding site, and the other feeding site is set empty.

We represent the food amount of the first feeding site and the one of the new feeding site as $f_1(t)$ and $f_2(t)$, respectively.
The initial food amount at the first site, $f_1(0)$, was set enough to contrast the stable pheromone trail between the first food source site and the nest site.
We set $f_1(0)=20$ herein.
The time evolution of the amount of food obeys:
\begin{subequations}
\begin{align}
	f_1(t+\tau)&=f_1(t) - n_1^f(t) f_e, \\
	f_2(t+\tau)&= \begin{cases} 0, & t < t_s \\
	20, & t=t_s \\
	f_2(t) - n_2^f(t) f_e, & t_s < t 
 \end{cases}
\end{align}
\end{subequations}
where $f_1(0) = 20$, and $t_s$ is the time when the first food source became empty.


\section{Evaluation of the Colony Ability} 
In this section, we propose three types of index used to evaluate the foraging-related abilities of individual colonies in our simulations: foraging efficiency $F$, transportation efficiency $E_T$, and discovery efficiency $E_D$.

\subsection{Foraging Efficiency}
First, we define the foraging efficiency $F$ of a colony: the average rate that ants return to the nest with food.
The continuous feeding method introduced in the previous section is employed to calculate this quantity.
We calculate the foraging efficiency $F$ as a function of $\hat{N}_n$ and $\alpha_{s}$ defined as follows:

\begin{equation}
\label{eq:effect}
F(\hat{N}_n, \alpha_{s}; \theta, \delta) = \frac{1}{T}\sum_{t=0}^{T}n_h(t),
\end{equation}
where $n_h(t)$ is the number of ants reaching the nest carrying food from one of the food sources.

\begin{figure*}[tp]
\centering
\includegraphics[width=\textwidth]{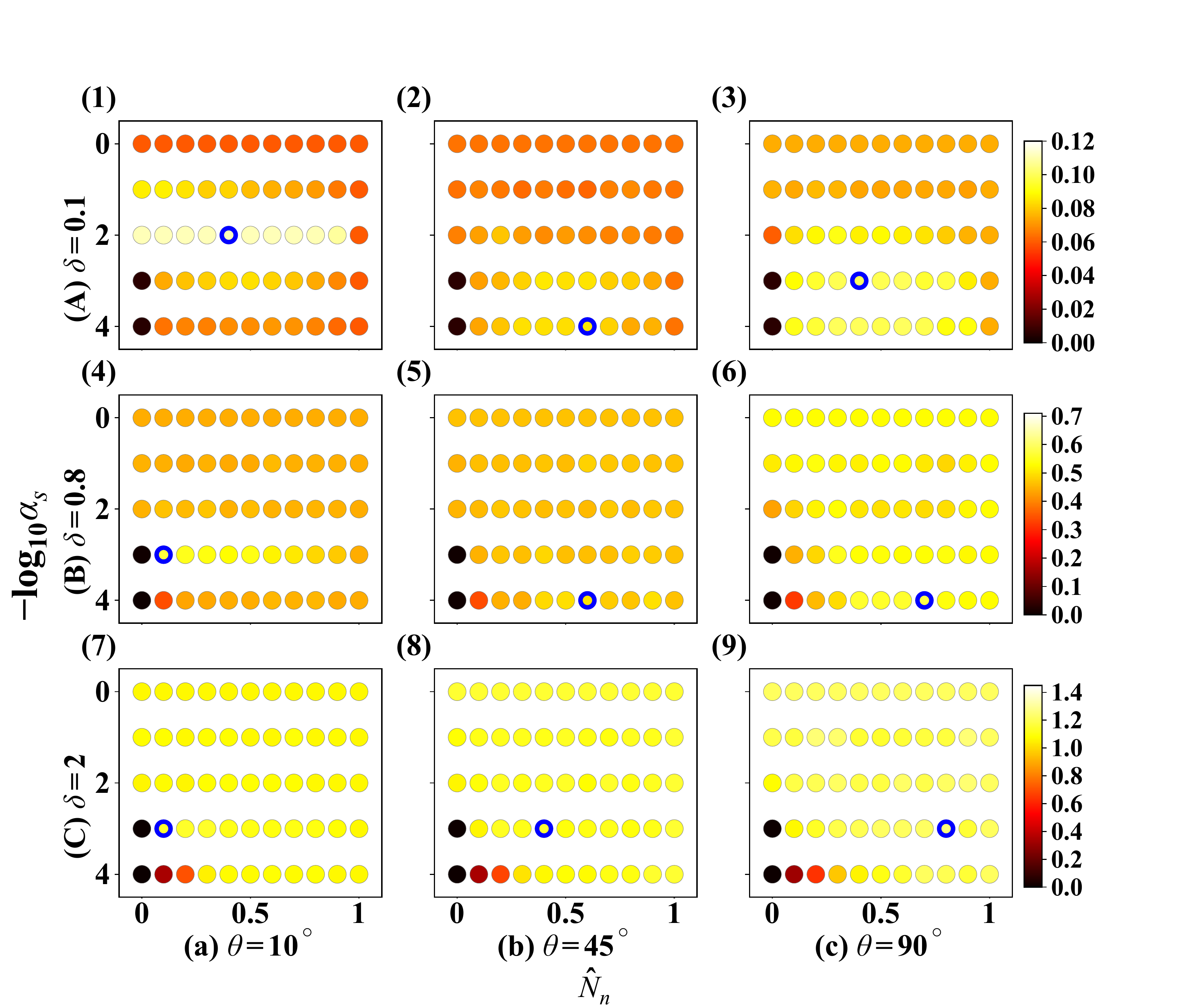}

\caption{Landscape of the averaged foraging efficiency, $F(\hat{N}_n, \alpha_{s}; \theta, \delta)$, where the total number of ants in a colony is fixed as $N_{total}=500$. The points enclosed by bold blue lines represent the maximum point in each sub-figure.}
\label{pd_F}
\end{figure*}

\begin{figure*}
	\includegraphics[width=\textwidth]{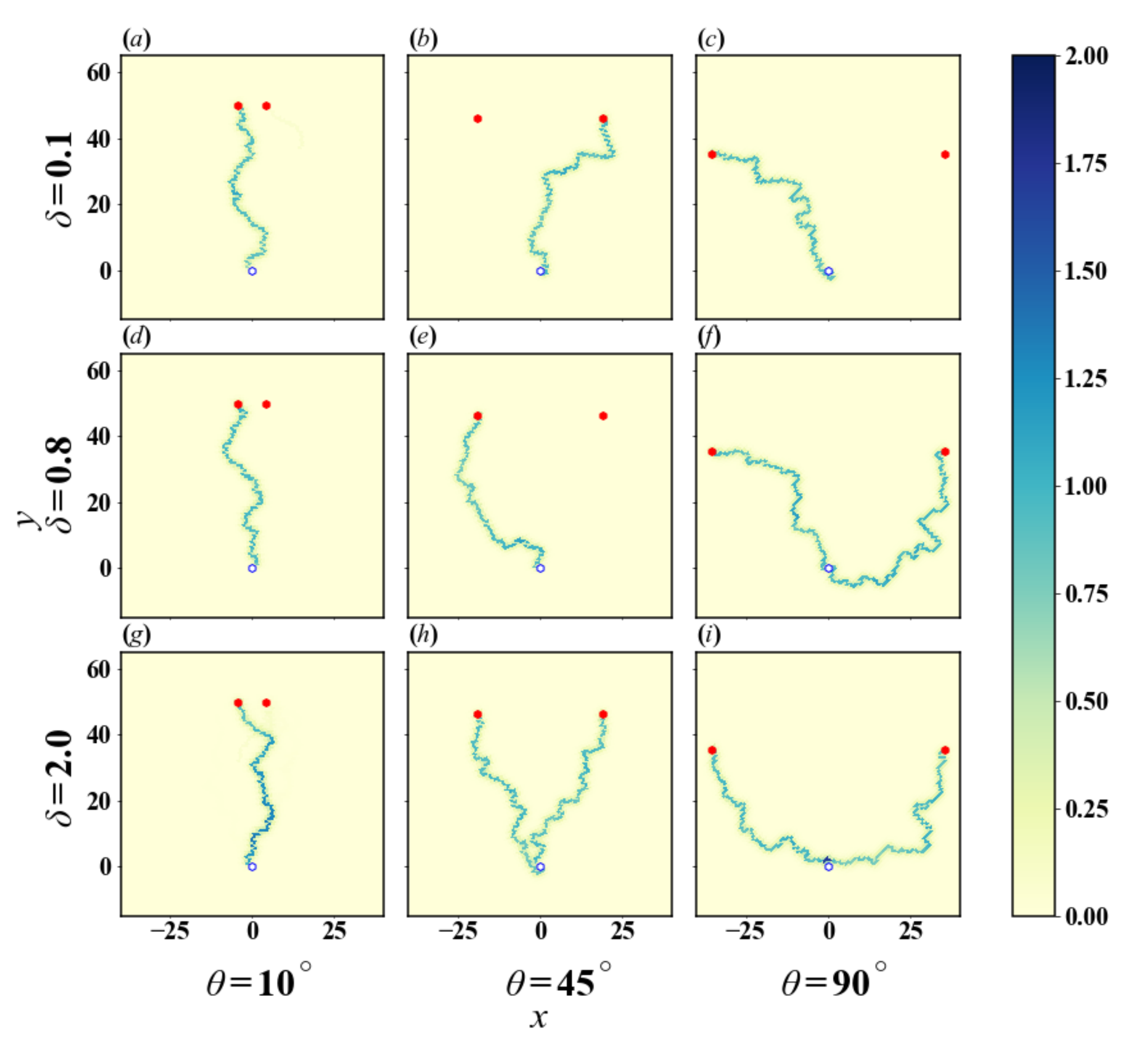}
	\caption{\label{fig:lastphero}Sample snapshots of the pheromone trails for different parameters at the final state. (a)(d)(g)$-\log\alpha_s=3, \hat{N}_n=0.1$ (b)(e)(h)$-\log\alpha_s=4, \hat{N}_n=0.5$ (c)(f)(i)$-\log\alpha_s=4, \hat{N}_n=0.4$}
\end{figure*}

We observed the ``foraging efficiency landscape" $F(\hat{N}_n, \alpha_{s}; \theta, \delta)$ on the two parameters, $\hat{N}_n$ and  $\alpha_{s}$, spaced at each pair of the relative angle $\theta$ between the two food resource sites (\figref{FoodSources}) and the food increment $\delta$ in Eq. \eqref{eq:food}.
The distance from the nest to food sources $R$ is fixed as $R=50$.
All calculations start from $t=0$, at which all ants are located in the nest and last until $T=10000$, and the statistical results are averaged over 100 ensembles.
This duration of calculation is sufficiently longer than the characteristic time for the formation of pheromone trails. 

\begin{figure*}[tp]
	\includegraphics[width=1.1\textwidth]{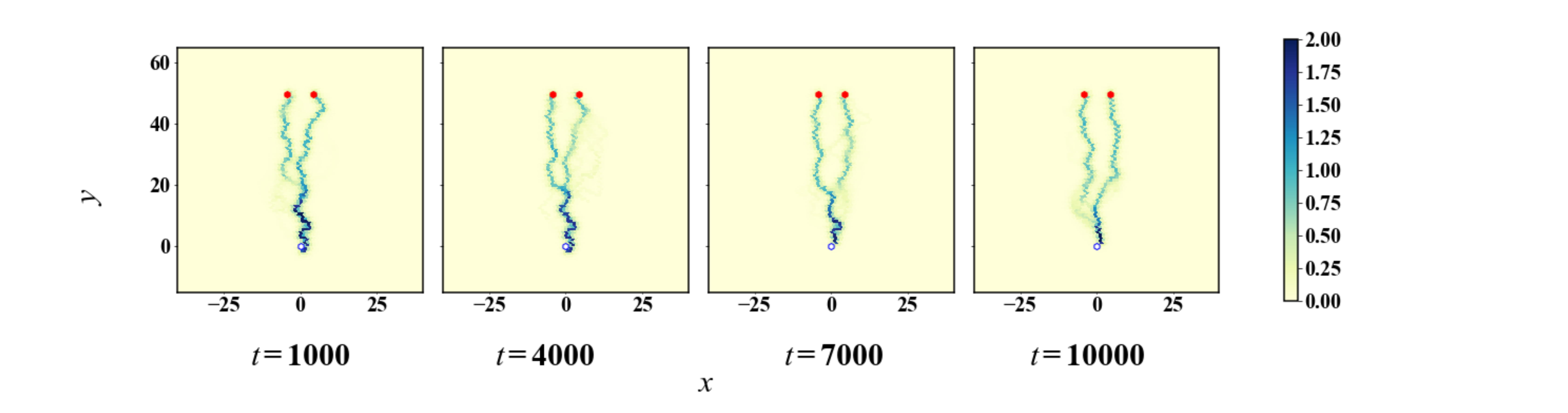}
	\caption{\label{fig:timedev}Time development of the pheromone trail on the field. Red hexagons are the food sources, while the white hexagon is a nest. $\theta=10^\circ$ $\delta=0.1$, $-\log_{10}\alpha=3$, $\hat{N}_n=0.1$.}
\end{figure*}

\subsection{Transportation and Discovery Efficiencies}

The foraging efficiency $F$ is measured as an averaged quantity over a sufficiently longer time than the characteristic time for trial formation connecting a nest and the food source sites with the simulations assuming a stationary environment.
We investigate the short-term dynamical aspects of the foraging activity through two kinds of quantities to elucidate the basic mechanism and generate the characteristic features of the landscape of $F$: $E_T$, the transportation efficiency and $E_D$, the discovery efficiency.
We employ the specialized feeding method to calculate $T_T$, which is the
time taken for the second food in each simulation to run out since it is found by an ant, and $T_D$, which is the duration between the timing of the first food running out and the second food found by an ant.
Note that we define the timing of the second food found as the time the amount of the second food is reduced to $90\%$ of the initial amount to distinguish the random access and collective foraging.
Using $T_T$ and $T_D$, we quantify the transportation efficiency of food $E_T=1/T_T$ and the discovery efficiency of food $E_D=1/T_D$.
With respect to $T_D$, a shorter duration means faster finding of new food at the alternative feeding site and high discovery efficiency $E_D$.
The two kinds of efficiencies $E_T$ and $E_D$ are investigated after the simulations to measure the landscape of $F$ for each set of $(\theta, \delta)$ are performed.
Therefore, these efficiencies are represented as a function of $\hat{N}_n$, $\alpha,$ and $\theta$: $E_T=E_T(\hat{N}_n, \alpha_{s}; \theta)$ and $E_D=E_D(\hat{N}_n, \alpha_{s}; \theta)$.

\section{Numerical Results}
We build the landscapes for the quantities $F(\hat{N}_n, \alpha; \theta, \delta)$ with the continuous feeding method and the ones for  $E_T(\hat{N}_n, \alpha; \theta)$ and $E_D(\hat{N}_n, \alpha; \theta)$ with the specialized feeding method introduced in the previous section by varying the system parameters $\hat{N}_n$ and $\alpha_s$.
First, we calculate the landscape of the foraging efficiency $F(\hat{N}_n, \alpha; \theta, \delta)$ to analyze what type of combination of $\alpha_s$ and $\hat{N}_n$ is optimal to achieve the most efficient foraging with each combination of the parameters of the continuous feeding method, $\theta$ and $\delta$.
Thereafter, the transportation efficiency and the discovery efficiency are calculated to analyze the short-term dynamics by varying $\alpha_s$ and $\hat{N}_{n}$.
With respect to the angle between the two food sources $\theta$, it can be $ 0 < \theta \le 180^\circ$; however, we discuss the results for $\theta \le 90^\circ$ in this report because the simulation results of the cases, where $\theta > 90^\circ$, show that the all efficiencies qualitatively depicted similar landscapes to the case $\theta = 90^\circ$.
Thus, we focus on the cases $\theta \le 90^\circ$.
As for the feeding environment, we conduct simulations using the continuous feeding method for three values of $\delta$; $\delta=0.1$ as a starving feeding environment, $\delta=0.8$ as an intermediately feeding environment, and $\delta=2.0$ as a saturated feeding environment, in which the amount of food is always at or around the upper limit $f_c$.

\begin{table}[tb]
    \centering
    \caption{Optimal combinations of $(\hat{N}_n, \alpha_s)$, which achieved optimal $F(\hat{N}_n, \alpha_s;\theta, \delta)$.}
    \label{tab:opteff}
    \begin{tabular}{c c c c}
	    \hline
        \toprule
        \midrule
            \multirow{2}{*}{\hspace{5mm}$\delta$\hspace{5mm}}& \multicolumn{3}{c}{$\theta$}\\
            \cmidrule(lr){2-4}
            & \hspace{5mm} $10^\circ$ \hspace{5mm} & \hspace{5mm}$45^\circ$ \hspace{5mm}& \hspace{5mm}$90^\circ$\hspace{5mm} \\
		    \cmidrule(r){1-1}\cmidrule(l){2-4}
             \hline
            \multicolumn{1}{c}{0.1}& $(0.4, 10^{-2})$ & $(0.6, 10^{-4})$ & $(0.3, 10^{-4})$ \\
            \multicolumn{1}{c}{0.8}& $(0.1, 10^{-3})$ & $(0.6, 10^{-4})$ & $(0.8, 10^{-4})$ \\                                                                                                                                                                                                                                                                                                                                                                                                                                                                                                                                                                                                                                                                                                                                                                                                                                                                                                                                                                                                                                                                                                                                                                                                                                                                                                                                                                                                                                                                                                                                                                                                                                                                                                                                                                                                                                                                                                                                                                                                                                                                                                                                                                                                                                                                                                                                                                                                                                                                                                                                               
            \multicolumn{1}{c}{2.0}& $(0.2, 10^{-3})$ & $(0.4, 10^{-3})$ & $(0.8, 10^{-3})$ \\
        \midrule
        \bottomrule
    \hline
    \end{tabular}
\end{table}

\subsection{Foraging Efficiency for Two Food Sources}
Figure \ref{pd_F} shows the landscapes of the averaged foraging efficiency $F(\hat{N}_n, \alpha_{s}; \theta, \delta)$ for the different continuous feeding environment parameters, $\theta$ and $\delta$.
The figure shows that the optimal combination of $(\hat{N}_n, \alpha_s)$ found neither along the line $\hat{N}_n=1$ nor along $\alpha_s=\alpha_n$ varies depending on $\theta$ and $\delta$.
In other words, some sort of stochasticity among ants in the colony is indispensable to achieve the optimal efficiency, and the form of the optimal distribution of the stochasticity is affected by the environment conditions.
Figure \ref{fig:lastphero} represents the pheromone trails at the final state of the simulation showing that the dynamical pheromone trails have many types of trails.

This result, as a whole, supports the previously introduced idea of `strategy of error.'
However, the following outcome listed on Table \ref{tab:opteff} supplies us with the extended concept of the previous one.
More specifically, in the case of $\delta=0.1$, the optimal distribution of the stochasticity largely depends on $\theta$.

Figure \ref{pd_F}(1), when $\theta=10^\circ$, shows that the efficiency is optimized when $\alpha_s=10^{-2}$ and a finite number of standard ants, $\hat{N}_n=0.4$, is mixed in the colony.
In these cases of $\delta=0.1$, the food at the sources easily runs out because of the slow food refill.
Consequently, these results suggest that, for an efficient foraging with continuous feeding method, ants have to find another feeding site when the food sources at the discovered site run out.
A finite number of weak stochastic ants are required to keep their mobility between alternative feeding sites, and concurrently, to remain enough transporting ability.
As shown in Fig. \ref{fig:timedev}, the pheromone trail dynamically changes when the $F(\hat{N}_n, \alpha_s; \theta, \delta)$ reaches maximum values: $\hat{N}_n=0.4, -\log\alpha_s=2, \theta=10$, and $\delta=0.1$.
We discuss these hypotheses in the next sub-section.

As for the others, $\theta=45^\circ, 90^\circ$, the landscapes are different from that for $\theta=10^\circ$.
Figures \ref{pd_F}(2) and \ref{pd_F}(3) show that the landscapes are similar to each other, and the vicinity of the maximum value is relatively flat.
Compared with the landscape of $\theta=10^\circ$, the parameters that represent the maximum $F$ shift to the region where the sensitivity strength and the ratio of stochastic ants are higher.
In the case of $\theta=45^\circ, 90^\circ$, to achieve the optimized foraging, stronger stochastic ants are required, and a finite number of normal ants have to exist in the colony.

In the case of $\delta=0.8$ (\figref{pd_F}(4)--(6)), the trends of the deformation of the landscapes of $F(\hat{N}_n, \alpha_s; \theta, \delta)$ under the change of $\theta$ are kept similar with the abovementioned case of $\delta=0.1$; however, the sharpness of the landscapes around the optimized combinations of $(\hat{N}_n, \alpha_s)$ is not very clear as the abovementioned case.
A possible reason for this is that the food sources do not easily run out compared to the above.
Thus, the ants that find one of the food sources can easily obtain a finite amount of food even if they fail to find the alternative food source.
This hypothesis is also discussed in the next sub-section.

In the case of $\delta=2.0$ (\figref{pd_F}(7)--(9)), the food do not run out even if all normal ants transport the food to the nest.
The existence of the stochastic ants with a very strong stochasticity does not contribute the optimal foraging in any value of $\theta$ because the stochastic ants only reduce the opportunity to transport food.

Moreover, for insufficient food supply ($\delta=0.1$ and $0.8$), the maximum $F$ is the lowest at $\theta=45^\circ$.
This phenomenon is due to the interference to exploratory behavior by the constructed pheromone path to achieve effective exploitation.
In the case $\theta =45^\circ$, the interference by the already established path to one food source decreases the chance for normal ants to construct new path to the other food source by following the pheromone secreted by the strongly stochastic ants.
Here, the “interference” means that the already established path to one food may, sometimes, disturb the construction of new path because it certainly attracts the normal ants.



\subsection{Transportation and Discovery Efficiencies}

\begin{figure*}[tb]
\centering
\includegraphics[width=\textwidth]{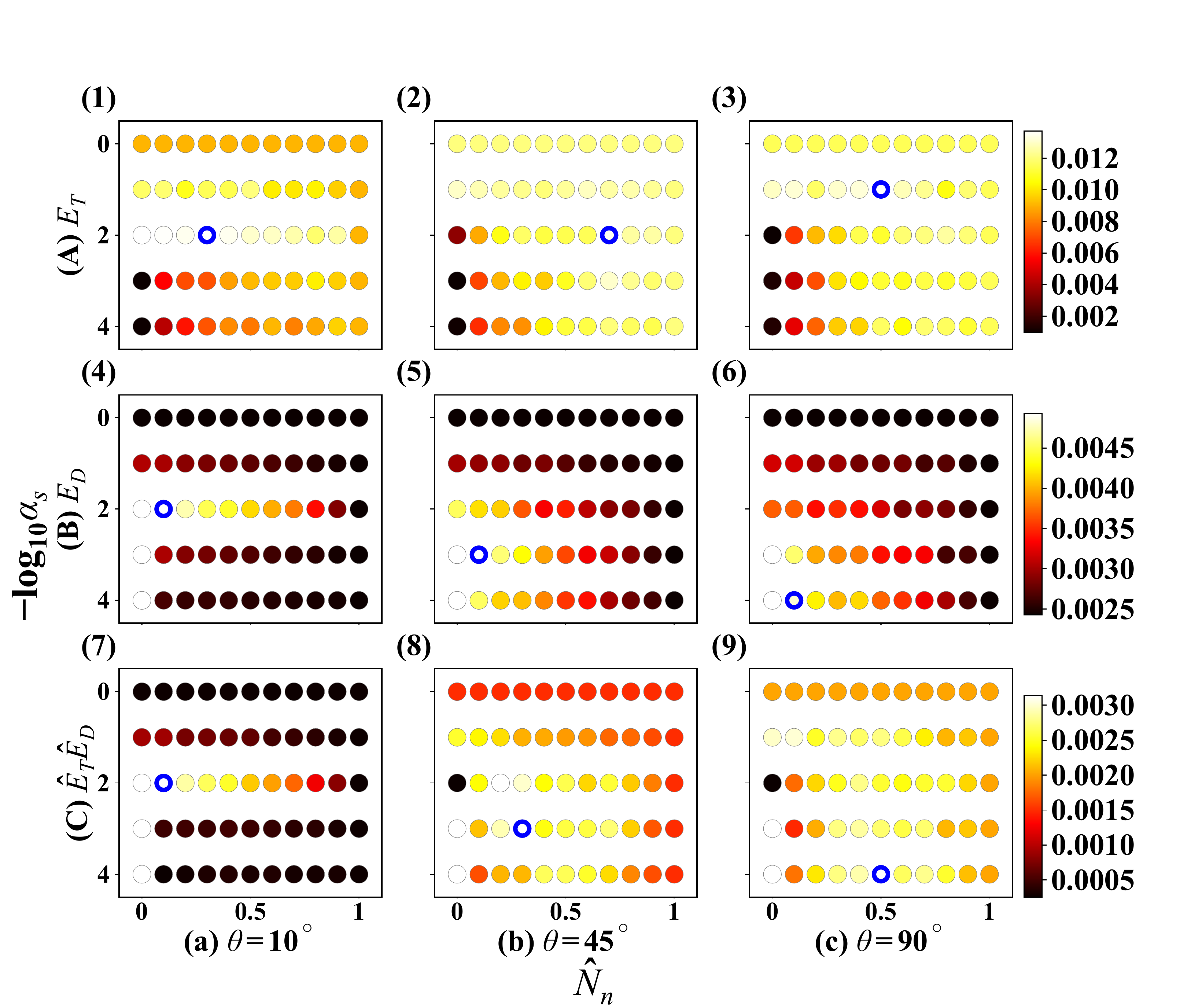}
\caption{Landscape of the transportation and discovery efficiencies, $E_T(\hat{N}_n, \alpha_{s}; \theta)$, $E_D(\hat{N}_n, \alpha_{s}; \theta)$, and $\hat{E}_T\hat{E}_D$, for different angles between food, $\theta$, where the total number of ants in a colony is fixed as $N_{total}=500$. The points enclosed by the heavy blue lines represent the minimum point in each sub-figure, except for the extreme values, $\hat{N}_n=0$ or $-\log\alpha_s > 2$.}
\label{CharaTimes}
\end{figure*}

The previous section presented the landscape of $F(\hat{N}_n, \alpha_{s}; \theta, \delta)$ with the continuous feeding method and confirmed that the stochasticity in following pheromone trails increased the foraging efficiency of ants in a colony, assuming that the two food sources were relatively close.
This result supports the basic idea of the `strategy of errors' by Deneubourg\cite{De0}.
Especially, in the case, where two food sources are separated over a certain distance, $\theta=90^\circ$ and the amount of food is not sufficient $\delta=0.1$ or $0.8$, the contrasted coexistence of highly and weakly stochastic ants is the task allocation on errors for the optimal foraging, which seems to supply us with the extended concept of the strategy of error.
However, the mechanism that improves the foraging efficiency in the latter case is still unclear.
Thus, we discuss herein the short-term dynamics in the foraging behaviors using the specialized feeding method to understand how $\hat{N}_n$ and $\alpha_s$ affect the collective dynamics by analyzing $E_T$ and $E_D$ introduced in the previous section, as shown in \figref{CharaTimes} which presents the numerical results.
$E_T$ represents the ability of ants to transport the food source to the nest, while $E_D$ characterizes the ability of ants to promptly explore the alternative food source resisting the attraction of the remaining pheromone around the initially found food source.
$E_T$ and $E_D$ are interpreted as the parameters characterizing the abilities of ``exploitation" and ``exploration", respectively.
The required degree of abilities largely depends on the distance between the food sources on $\theta$.

In the case where $\theta=10^\circ$, the landscapes of $E_T$ and $E_D$ have similar optimal combinations $(\hat{N}_n, \alpha_s)$ to that of $F$.
In other words, the existence of mildly stochastic ants in the colony improves the efficiencies in short-term dynamics.
At the same time, the foraging efficiency in the long term is improved with the optimal parameters.
In contrast, in the case where $\theta=45^\circ$ and $90^\circ$ (\figref{CharaTimes}(2--3) and (5--6)), the landscapes of $E_T$ and $E_D$ do not have optimal parameters similar to those of $F$.
Note that in the cases where parameters $\hat{N}_n=0$ and $-\log \alpha_s > 2$ (\figref{CharaTimes}(4--6)), $E_D$ shows extremely higher values than the other combinations of parameters because the strong stochasticity prevents colonies from constructing pheromone trails, and the spatial symmetry is not broken, indicating that no pheromone trail is constructed in the simulation.
Accordingly, the extreme values $\hat{N}_n=0$ or $-\log\alpha_s > 2$ are ignored in the following discussion on the effect of the optimal combination of the stochasticity.

The landscapes of $\hat{E}_T\hat{E}_D$, where $\hat{E}_T = E_T / \smash{\displaystyle\max_{(\hat{N}_n, \alpha_s)}} E_T$ and $\hat{E}_D = E_D / \smash{\displaystyle\max_{(\hat{N}_n, \alpha_s)}} E_D$, are similar to the ones of $F$, including the case where $\theta=10^\circ$ (Table \ref{tab:opteff} and \ref{tab:opttime}), except for the abovementioned extreme values of $(\hat{N}_n, \alpha_s)$.
These results indicate that the mechanism for achieving optimal foraging in the colony with the insufficient feeding environment can be discussed based on the short-term dynamics using the specialized feeding method such that the colony of ants must be equipped with the balance of abilities of finding a new food source and efficiently carrying food to the nest using the pheromone trail.
The product, $\hat{E}_T\hat{E}_D$, represents the possible function to characterize the feature of ``exploitation--exploration" trade-off in this foraging system.

\begin{table}[bt]
    \centering
    \caption{Optimal combinations of $(\hat{N}_n, \alpha_s)$, where $E_T$, $E_D$, and $\hat{E}_T\hat{E}_D$ are optimal for each.}
    \label{tab:opttime}
    \begin{tabular}{c c c c}
    \hline
        \toprule
        \midrule
            \multirow{2}{*}{}& \multicolumn{3}{c}{$\theta$}\\
            \cmidrule{2-4}
            & \hspace{5mm}$10^\circ$ \hspace{5mm} & \hspace{5mm}$45^\circ$ \hspace{5mm}& \hspace{5mm}$90^\circ$\hspace{5mm} \\
		    \cmidrule(r){1-1}\cmidrule(l){2-4}
            \hline
            \multicolumn{1}{c}{$E_T$}& $(0.3, 10^{-2})$ & $(0.7, 10^{-2})$ & $(0.5, 10^{-1})$ \\
            \multicolumn{1}{c}{$E_D$}& $(0.1, 10^{-2})$ & $(0.1, 10^{-3})$ & $(0.1, 10^{-4})$ \\
            \multicolumn{1}{c}{$\hat{E}_T\hat{E}_D$}& $(0.1, 10^{-2})$ & $(0.3, 10^{-3})$ & $(0.5, 10^{-4})$ \\
        \midrule
        \bottomrule
    \hline
    \end{tabular}
\end{table}

\section{Summary}
In summary, through the simulations of a mathematical model reflecting the behavioral parameters of the existing species of ants, we confirmed herein that the coexistence of highly stochastic ants and normal or weakly stochastic ants in the same colony is the key to achieving the optimal foraging behavior in an environment with insufficient food supply at finitely separated food sources.
In an environment with a sufficient feeding environment or in the case when two feeding sites are located nearby, the feeding efficiency was maximized when the colony was composed of only the weakly stochastic ants or their mixture with normal ants.
In the latter case, stable food sources can be achieved by continuing to access the same or nearby source through the shortest pathway; however, in this case, the highly stochastic ants cannot follow the pheromone pathway, decreasing their ability to carry food. In contrast, mildly stochastic ants prevent the colony from being trapped with a too attractive, but non-optimized way of trail.
In the former case, the food at each food source was easily and quickly exhausted; thus, the strategy relying on a single food source increased the risk of losing the source because of sudden accidents.
Therefore, the probability of discovering other food sources must be increased such that the colony does not lose accessible food sources.
In other words, the reason for the coexistence of highly stochastic and weakly stochastic ants in the same colony is such that the colony can adapt to unreliable external environments.
We call this effect the ``diverse-stochasticity-induced optimization," which is a type of role division of ants based on the degree of stochasticity to complete tasks with the highest efficiency.

Our result is consistent with the famous ``explore--exploit" problem, which has been discussed in many previous studies \cite{Hills46-54} in the theoretical and experimental viewpoints.
Gordon\cite{Gordon2017} reported the observation results that ants, which do not follow the established trails, could find better routes and other food sources in nature.
The existence of stochasticity in a multi-agent system is a main concern, but the quantitative difference in the strength of stochasticity is omitted.
Our result elucidate that the strength of stochasticity in ant behavior could be an intrinsic and important parameter in the multi-agent system.
The concept of diverse-stochasticity-induced optimization appears to be applicable to a class of systems wider than that in the present case.
However, the details depend on the situation of individual systems.
For example, a more complex environment like multiple-food source existence\cite{Latty2011} have different space and time scales for each food source.
Our results suggest that the optimal foraging with appropriate stochasticity parameters could be achieved.

An ant colony has many types of workers.
Some of which protect the queen, while others care for the larvae.
Our research suggested another class of role division, that is, a role division based on the degree of stochasticity regarding their behavior.
\acknowledgements
This study was partially supported by JST CREST grant number JPMJCR15D4(MS and HN) and JSPS KAKENHI grant numbers 16H04035(HN) and 26610117(HN).
We deeply appreciate the reviewers for their comments and suggestions that improved our study.

%
%
%
%
%
\end{document}